# Oscillations of a System of Coupled Oscillators with a Virtod-Type Virtual Cathode

A. E. Hramov

**Abstract**—The results of numerical modeling of oscillations in a system of two coupled oscillators operating on the basis of a virtual cathode (VC) with controlled external feedback are presented. It is shown that there exist the modes with generation of chaotic signals and synchronization modes with high signal-to-noise ratio. The effect of parameters of the electron beam and external feedback on the generation frequency of the system and phase relations between the signals generated in each module is analyzed.

**1.** Microwave oscillators, in which an electron beam with the virtual cathode (VC) is used as an active medium—the so-called vircators [1]—are promising devices of relativistic electronics. The VC devices attract an interest of researchers first of all because they are characterized by simple design, high level of generated power (from hundreds of megawatt to thousands of gigawatt), and the possibility of controlling the device by an external microwave signal. The latter constitutes the main condition that allows one to apply vircators as units of phased arrays (see, for example, [2–4]). A classical scheme of sending the control signal to a vircator system is based on the assumption that the external signal influences directly upon the VC (see [3–5]). However, such an approach requires a high power level of the control signal, which hampers practical applications of such systems. In this way, we note an array module [6] operating on the basis of the virtod-type vircator [7]. In this case, an external signal that influences upon the electron beam in the acceleration domain leads to modulation of the flow that enters the drift space; according to experimental investigations of the modulation of high-current electron beams [8], the required level of the control signal can be decreased by an order of magnitude. At the same time, it is well known that a preliminary modulation of the electron beam strongly influences the VC dynamics [9]. In [9], I demonstrate the possibility of synchronizing oscillations in the diode gap with overcritical current and a preliminarily modulated electron beam at the modulation frequency; in this case, the phase difference between the control signal and the signal generated by the VC oscillations tends to zero, which is an optimal condition providing an increase in the output power of a multi-module system.

In this paper, I present the results of numerical investigation of oscillations in a system of two coupled virtod-type oscillators. Oscillation characteristics of the system in the synchronization mode are of primary interest; I also analyze the dependences of the frequency and phase relationships of oscillations in each module on the parameters of the external delayed feedback.

**2.** The physical system under consideration is simulated by a set of two identical vacuum diodes; a high-current electron beam with the relativistic factor $\gamma_0$ = 3.5 is injected into the system. The beam current is denoted by $I_1$ and $I_2$, respectively, for the first and second diodes. Before entering the interaction space of each virtod, the beam is modulated in a narrow gap by the electromagnetic signal taken from the interaction space of another virtod; this signal is sent to the modulator with a certain delay in time. I consider a reditron model of the vircator [10]; according to this model, the electrons reflected from the VC are completely fallen out on the anode, which prevents the accumulation of oscillating charged particles in the cathode–VC gap. Such a scheme enables us to increase the signal-to-noise ratio and efficiency of the energy transformation as compared with the classical vircator [11, 12].

Let us characterize the main parameters that govern the system behavior: the ratio $\alpha_i$ of the beam current $I_i$ to the limit vacuum current $I_0$ and parameters of the external feedback line that couples the moduli, namely, the delay time $\tau_i$ and the coupling coefficient $A_i$ that specifies the modulation depth $m$ of the electron beam at the exit of the modulator (at the entrance to the drift space; here, $i$ denotes the number of the virtod). In this paper, I assume, unless otherwise specified, that the coupling between moduli is symmetric; that is, $\tau_1 = \tau_2 = \tau$ and $A_1 = A_2 = A$. The coupling coefficient is chosen so that the modulation depth of the beam at the exit of the modulator would take the value $m \sim 15\%$.

Numerical modeling of nonlinear time-dependent processes in the system under study was performed using a standard method of "particles in a cell" (PIC-method) [13] which is usually applied for solving such problems.

I assume that a sufficiently strong magnetic field is applied along the longitudinal axis $z$, so that the beam electrons are completely magnetized and one may consider one-dimensional motion of electrons. In each module, the beam is represented as a set of charged particles that are injected to the interaction space in equal time intervals. The dynamics of each large particle is determined from the solution of the relativistic equations of motion

$$\frac{\partial u_k}{\partial t} = \frac{Q_k}{m_k}E(z_k), \quad \frac{\partial z_k}{\partial t} = \frac{m_k c^2}{W_k}u_k, \quad k = 1, \ldots N, \quad (1)$$

performed at each time step. Here, $z_k$, $Q_k$, and $m_k$ are, respectively, the coordinate, charge, and rest mass of the $k$th particle; the speed of the particle $v_k$ is coupled with quantity $u_k$ by the relationship $v_k = u_k(1 + u_k^2/c^2)^{-1/2}$; $W_k = m_k c^2(1 + u_k^2/c^2)^{1/2}$; $E(z_k)$ is the field of the space charge at the point with coordinate $z_k$; and $N$ is the total number of particles. The field $E = -\partial\varphi/\partial z$ is determined, in the potential approximation, from the solution of the Poisson equation for potential $\varphi$ with the zero boundary conditions,

$$\frac{\partial^2\varphi}{\partial z^2} = \frac{\rho(z)}{\varepsilon_0}. \quad (2)$$

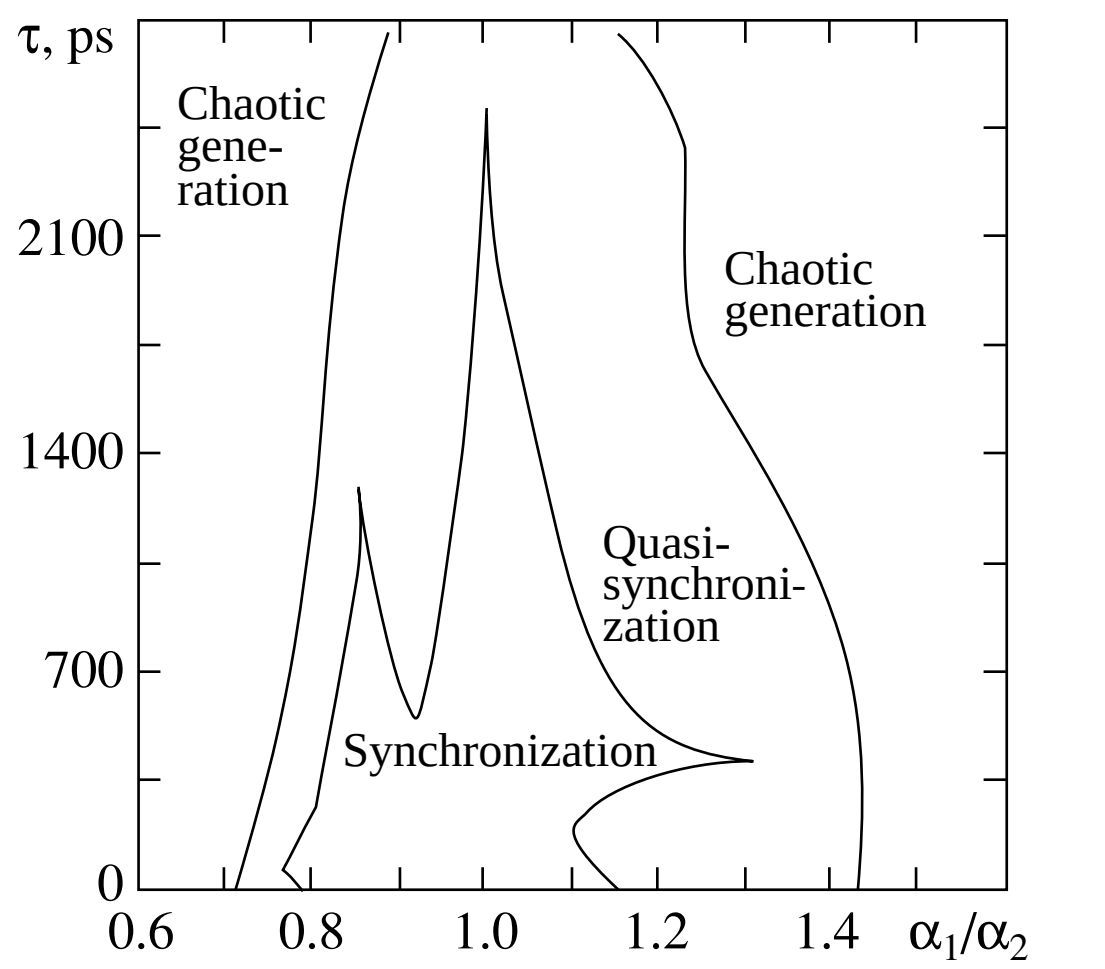

**Fig. 1.** The map of the oscillation modes of the system on the plane of parameters ($\alpha$ ,$\tau$).

Here, $\varepsilon_0$ is the dielectric constant. The current density of the space charge $\rho(z)$ is determined using the bilinear weighing at the nodes of the space grid,

$$\rho(Z_j) = \sum_{k=1}^{N} Q_k S(Z_j - z_k), \quad (3)$$

where $Z_j = j\Delta z$ is the coordinate of the $j$th node of the space grid, $\Delta z$ is its step, $S$ is a bilinear function that approximates the form of the particle,

$$S(z) = \begin{cases} 1 - |z|\Delta z, & |z| \le \Delta z. \\ 0, & |z| > \Delta z. \end{cases}$$

Equation of motion (1) was solved by the step-over method [13]; the Poisson equation (2) in the finite-difference form was integrated by the elimination method [14].

**3.** According to the results of numerical experiment, a system of coupled virtods demonstrates different operation modes depending on the parameters of the feedback line. When coupling coefficient $A$ corresponds to $m < 10\%$, the coupling is negligible, and oscillations in each virtod are almost autonomous. As the coupling coefficient increases, the mode of chaotic generation and the mode of synchronization with the high signal-to-noise ratio in the spectrum are observed depending on the frequency mismatch of the oscillator. The mode of synchronization takes place when $\alpha_1$ and $\alpha_2$ are close (the frequency mismatch $\Delta f = |f_2 - f_1|$ is small, where $f_1$ and $f_2$ are the oscillation frequencies in the autonomous modes corresponding, respectively, to $\alpha_1$ and $\alpha_2$) and the modulation depth $m \sim 10$–20%.

Figure 1 shows the zones of synchronization, quasi-synchronization, and chaotic generation in the coupled system constructed for $m = 15\%$ on the plane of parameters $\tau$ and $\alpha_1/\alpha$.

We see that when the frequency mismatch of the oscillators is small ($\Delta\alpha < 1$), the mode of synchronization in the system is observed when the signals of both oscillators are regular and have the same frequency; the phase difference between these signals takes a certain fixed value $\Delta\phi$ that does not vary in time. When the coupling between the moduli is symmetric, the synchronization domain on the plane of parameters gets narrow as the delay time increases.

The results of numerical experiment have shown that the width of the synchronization strip $\Delta f$ strongly depends on the value of coupling coefficient $A$. In the range of $A$ that corresponds to the modulation depth $m \sim 10$–15%, the width of the synchronization strip $\Delta f$ can be estimated as

$$\Delta f \sim \frac{\text{const}}{A^q} H(\tau - \tau_{\text{cr}}(A, \Delta\alpha)). \quad (4)$$

Here, $H(x)$ is the Heaviside function, $q \approx 1/2$, and $\Delta\alpha < 1$. The behavior of the system is characterized, in this case, by the presence of the threshold delay time $\tau_{\text{cr}}$; when this value is exceeded, the mode of synchronization is destroyed. Note that the critical value $\tau_{\text{cr}}$ of the delay time decreases when coupling coefficient $A$ and mismatch $\Delta\alpha$ increase.

For a large mismatch $\Delta\alpha \sim 1$, empirical dependence (4) is not valid, since, in this case, there exists the threshold time $\tau_{\text{cr, min}}$ that limits the delay time of the signal, at which regular oscillations in the system are observed, not only from above, but also from below. In the range $\tau_{cr,\,min} < \tau < \tau_{cr}$, the required value of the coupling coefficient that provides synchronization of oscillations must be sufficiently large ($m \sim 20\%$).

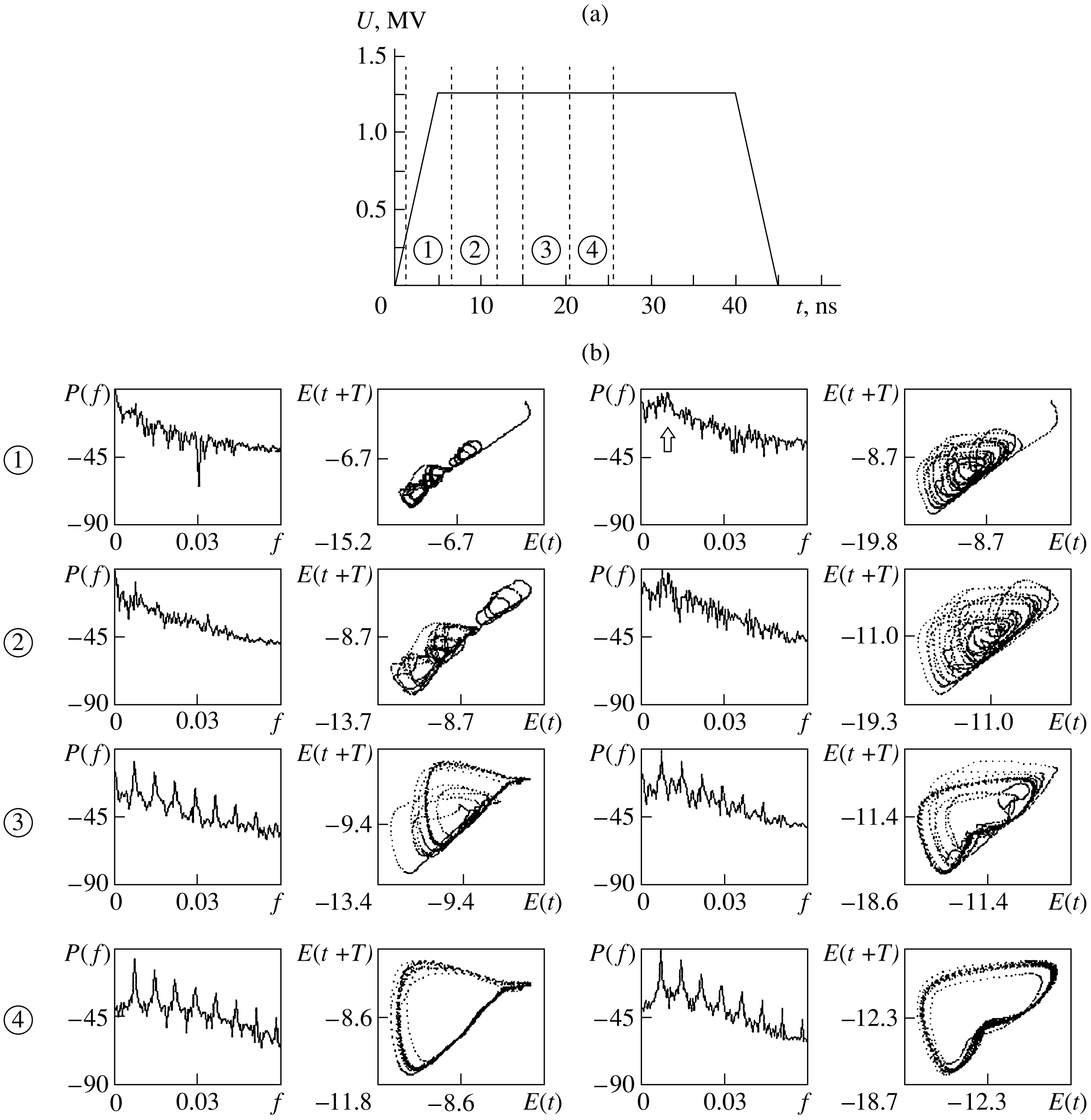

**Fig. 2.** Characteristics of oscillations in each oscillator determined at different moments of time during the action of the pulse of accelerating voltage.

Figure 2 shows the characteristics of oscillations in each oscillator determined at different moments of time during the action of the pulse of accelerating voltage sent to the diodes without any time shift. Figure 2a displays the pulse shape and the time intervals corresponding to the power spectra and projections of the phase portraits of the electric-field oscillations in the first (two right columns) and second (two left columns) diodes presented in Fig. 2b. The phase portraits were reconstructed by the method of Takens [15]. Note that the current of the second diode is greater than that of the first diode.

Figure 2 illustrates the process of establishing the mode of synchronization in the coupled system. In the beginning of the pulse of accelerating voltage, the spectrum of the oscillation power in each virtod is close to the noise spectrum [Fig. 2, (1)]. However, electrostatic instability in the flow of the second diode with a large current develops faster than in the flow of the first diode. In the spectrum of the second oscillator depicted in Fig. 2, (1), one can follow a weakly expressed peak (denoted by the arrow) that occurs at the frequency of oscillations in the autonomous diode with the same current. This leads to a deep modulation of the flow injected to the first diode, in which the VC is not yet formed. The flow modulation in the first diode exerts a pronounced influence on the conditions of the formation of the VC in this diode [9]. In time [Fig. 2, (2)], when the VC formation in the first diode continues (and its oscillation frequency is close to the generation frequency of the second diode due to the preliminarily modulation of the flow), the intensity of influence on the flow in the second diode increases. As a result, the generation frequency of both moduli is tuned to the mean frequency. The mean frequency is close to the frequency of autonomous oscillations in the second diode with a large current, which is caused by the aforementioned processes that occur at the initial stage. As the time increases further, the signal-to-noise ratio in the spectrum of the generation power of the system rises, the synchronization frequency and its harmonics are clearly distinguished, and the noise pedestal drops to the level of –45 dB in the domain of the first harmonic of the synchronization frequency and rapidly decays as the frequency increases [Fig. 2, (3, 4)].

The duration of the transition process that leads to the establishment of regular oscillations strongly depends on parameters of the feedback line. If mismatch $\Delta\alpha$ is sufficiently large, then, when $A$ increases, the duration of the transition process increases at small $\tau$ and decreases at large $\tau$. If the mismatch is small ($\Delta\alpha \ll 1$), the influence of parameters of a symmetric external feedback on the system behavior is much weaker; however, the introduction of an asymmetry to the system caused by unequal values of times $\tau_1$ and $\tau_2$ enables one to control the phase relationships between the signals in each module (see below).

Figure 3 illustrates the dependence of the generation frequency of the system on the ratio $\alpha_1/\alpha_2$ determined in the mode of synchronization for different $\tau$ and the generation frequency of the autonomous oscillator versus the beam current (curve *1*). Here, $\alpha_2 = 2.2$. As can be seen, the introduction to the system of the external delayed feedback yields a decrease in the generation frequency of the coupled system as compared with the autonomous diode. When the duration of the delay increases, the frequency decreases. The maximum readjustment of the generation frequency within the range of variation of $\tau$ is

$$(\Delta f/f)_{\max} \sim 20\%.$$

Considering the possibility of applying the system under study as a module of a phased array, one should note the importance of analysis of phase relationships between microwave signals in each virtod.

Figure 4 shows the phase difference $\Delta\phi$ between the signals in each virtod versus the delay time $\tau$ determined at the fixed $\alpha_1 = 2.2$. We see that, by varying the delay time $\tau$, one can readjust the phase difference in sufficiently broad limits; when the mismatch of oscillators is large, the influence of $\Delta\alpha$ on the phase difference is weak. When the mismatch of oscillators is small ($\Delta\alpha \ll 1$), the effect of the delay time of the symmetric coupling ($\tau_1 = \tau_2$) on the phase difference $\Delta\phi$ is negligible (curve *2*).

Introducing an asymmetry to the system connected with different delay times $\tau_1$ and $\tau_2$ ($\tau_1 \neq \tau_2$), one can control the value of $\Delta\phi$ in such a manner that the generation frequency remains almost unchanged. As can be seen from Fig. 5, a variation of the delay time $\tau_2$ at a fixed $\tau_1$ enables one to readjust the phase difference in sufficiently broad limits. For example, at $\tau_1 = 14$ ps, $\Delta\phi$ varies from 20 to 90° with respect to $\tau_2$. An increase in the fixed delay time $\tau_1$ deteriorates the monotonic dependence of $\Delta\phi$ with respect to increasing $\tau_2$. Note that when coupling coefficient $A$ (or the modulation depth of the flow injected to the drift space, which is the same) increases, the phase difference between oscilla-tions in both virtods decreases (see curves *1* and *3* in Fig. 5 that correspond to $m \approx 15$ and 25%).

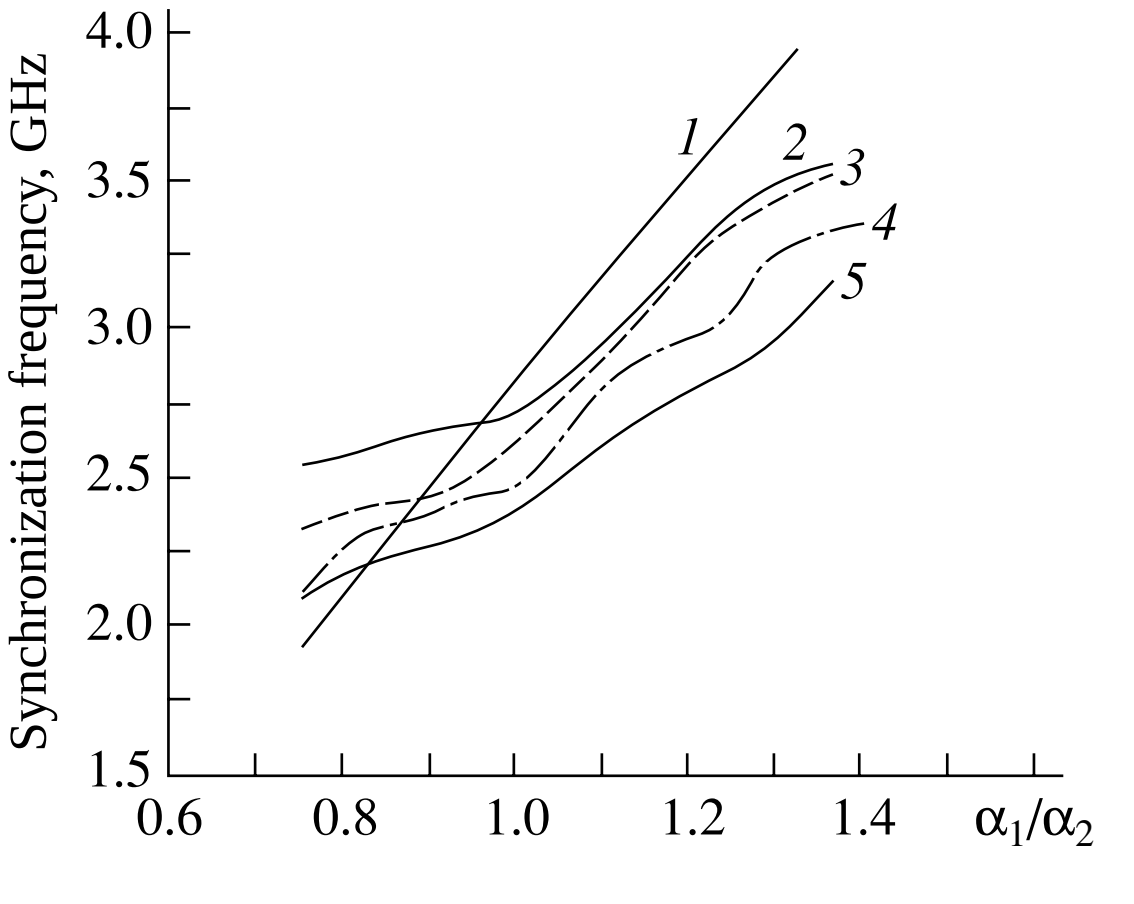

**Fig. 3.** The generation frequency of the system as a function of currents $\alpha_1/\alpha_2(\alpha_2 = 2.2)$. (*1*) autonomous oscillations; $\tau$ = (*2*) 14, (*3*) 413, (*4*) 825, and (*5*) 2.2 ns.

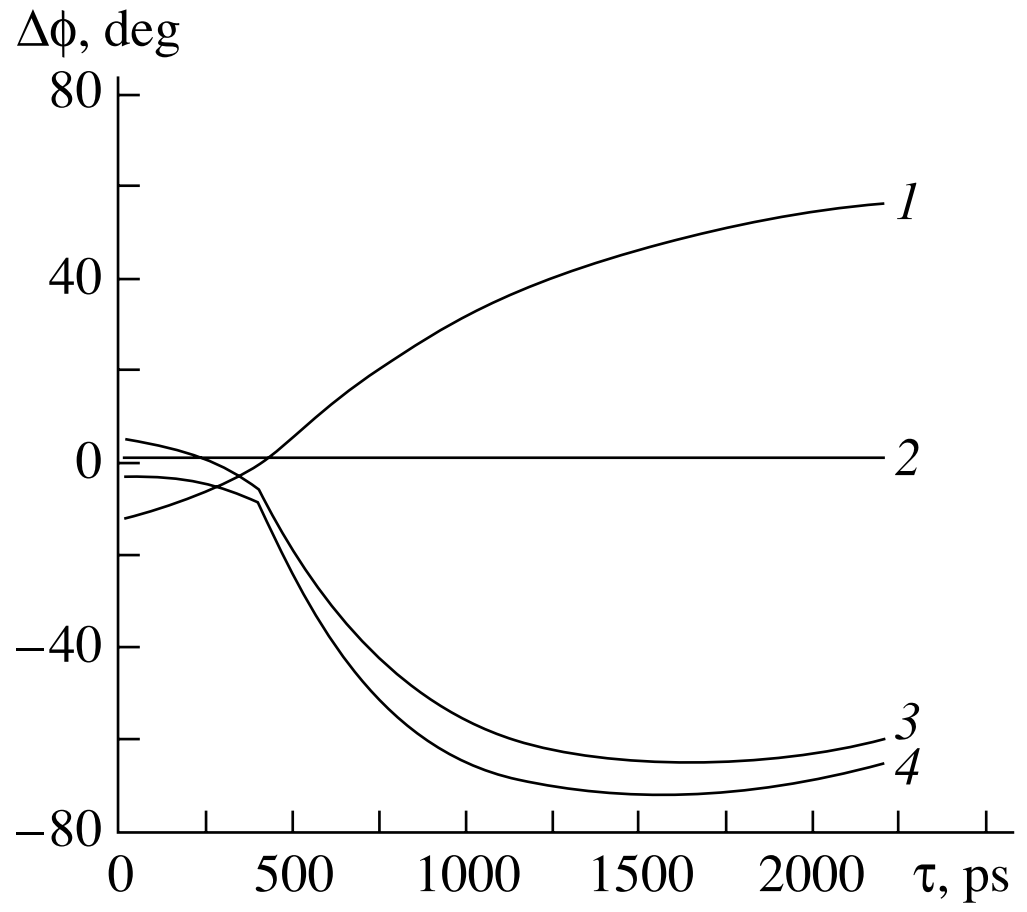

**Fig. 4.** The phase difference between the generated signals in each virtod vs. the delay time determined for different currents of the flows: $\alpha_1/\alpha_2$ = (*1*) 0.9, (*2*) 1.0, (*3*) 1.1, and (*4*) 1.2.

When the delay times $\tau_i > 2.5$ ns, the mode of synchronization is destroyed, and the system passes on the mode of chaotic oscillations through alternation [16]. In this case, the time realizations of the electric-field oscillations in each virtod have the form of periodic sequences broken by irregular splashes. As the delay time increases, the duration of the laminar phase of oscillations shortens.

In the domain of quasi-synchronization, the oscillations in each module occur at one definite frequency and, in this case, the mutual quasi-synchronization between the moduli takes place (that is, synchronization with a small difference between the frequencies of the signals generated in each module; the frequency difference depends on time and oscillates close to the zero mean value).

Chaotic oscillations are observed in the system when the mismatch of oscillators is large ($\Delta\alpha > 1$); these oscillations get complicated as the delay duration and mismatch $\Delta\alpha$ increase. In this case, there is a high noise pedestal in the spectrum of each virtod, which slowly decreases as the frequency increases, and weakly expressed peaks of the main frequency $f(\alpha_i)$ and its subharmonics are observed on the background of noise. The dimensionality $D$ of the reconstructed attractors calculated by the Grassberger–Procaccia algorithm [17] strongly depend, as well as the phase portraits, on the system parameters. The value of dimensionality $D$ is fractional, which is evidence of chaotic oscillations in the system. As the delay time increases, the dimensionality of the reconstructed attractors and, consequently, the oscillation complexity, increase; this effect is observed even when the mismatch of the oscillators is small (see Fig. 1, where the narrowing of the synchronization domain that occurs as $\tau$ increases is distinctly seen). When the coupling between the moduli is asymmetric ($\tau_1 \gg \tau_2$), a rigid transition to the mode of chaotic oscillations through alternation takes place. One may conclude that an increase in the delay time yields a complication of the system dynamics. Note that a similar effect is observed in the autonomous virtod. As is shown in [18], an increase in the delay time of the external delayed feedback yields a transition to the mode of chaotic oscillations in the electron beam with a VC. Thus, the above analysis enables one to consider the system under study operating in this mode as an oscillator of chaos that can be easily readjusted.

**Table**

| $\alpha_1/\alpha_2$ | $\tau$, ns | $D$ |
|---|---|---|
| 0.8 | 0.1 | 1.01 |
| | 0.5 | 1.52 |
| | 1.0 | 2.63 |
| 1.2 | 0.1 | 1.04 |
| | 0.5 | 1.11 |
| | 1.0 | 2.80 |
| 1.4 | 0.1 | 1.34 |
| | 0.5 | 1.85 |
| | 1.0 | 3.62 |

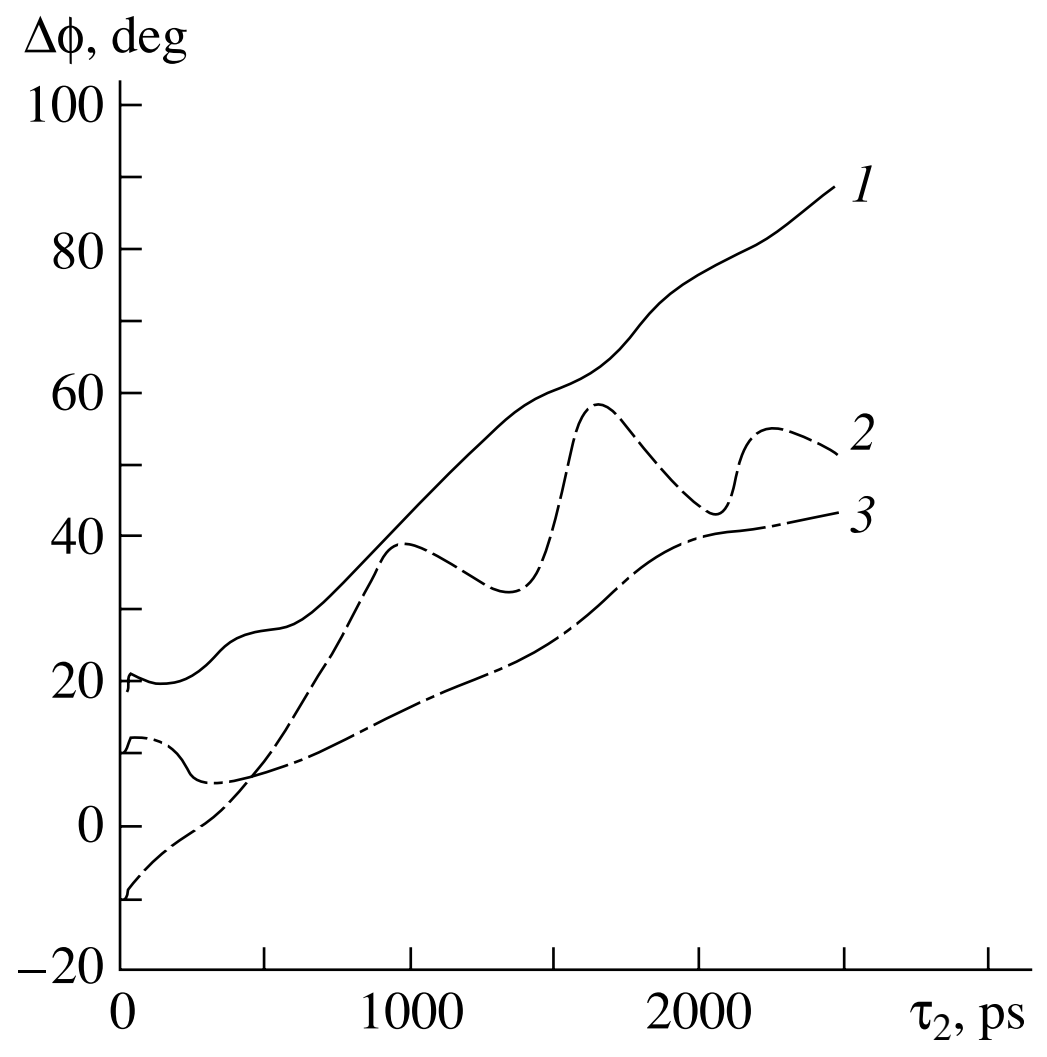

**Fig. 5.** Asymmetric case: the generation frequency of the system as a function of the delay time $\tau_2$ determined for different values of $\tau_1$ and $A$: (*1*) $\tau_1 = 14$ ps and $A = 2$ ($m = 15\%$); (*2*) $\tau_1 = 1.4$ ns and $A = 2$ ($m = 15\%$); and (*3*) $\tau_1 = 14$ ps and $A = 3$ ($m = 25\%$).

**4.** The study of the system of two coupled oscillators operating on the basis of virtod-type VCs has demonstrated the possibility of controlling the characteristics of such systems by varying the parameters of the external feedback. The dependence of both the frequency and generation frequency band on the delay time of the feedback line is observed. An asymmetric variation of the delay time in the mode of synchronization makes it possible to readjust the phase difference between the signals in each module within the limits 0–$\pi/2$. Thus, the results obtained enable one to consider the system under study as a component of systems with a larger number of the degrees of freedom, for example, pulse phased arrays operating on the basis of parallel vircators.

## ACKNOWLEDGMENTS

I am grateful to V.G. Anfinogentov for fruitful discussions and valuable remarks.

This work was supported by the Russian Foundation for Basic Research, project no. 96-02-16753.